\documentclass[preprint,5p,twocolumn]{elsarticle}

\journal{arXiv}
\usepackage{lineno}
\modulolinenumbers[5]

\usepackage[T1]{fontenc}
\usepackage[utf8]{inputenc}
\usepackage[dvipsnames]{xcolor}

\biboptions{sort&compress}

\usepackage{algorithm}
\usepackage{algpseudocode}
\usepackage{amsmath}
\usepackage{amsthm}
\usepackage{amssymb}
\usepackage{amscd}
\usepackage{bm}
\usepackage{booktabs}
\usepackage{latexsym}
\usepackage{mathtools}
\usepackage{comment}

\usepackage{graphicx}
\usepackage{epsfig}
\usepackage{epstopdf}
\usepackage{svg}
\usepackage[caption=false]{subfig}
\usepackage{placeins}
\usepackage{tabularx}

\usepackage{url}

\renewcommand{\vec}[1]{\boldsymbol{#1}}

\providecommand{\about}[0]{\raise.17ex\hbox{$\scriptstyle\sim$}}

\providecommand{\eqref}[1]{(\ref{#1})}

\providecommand{\vanish}[1]{}

\graphicspath{{figures}}

\begin{document}

\begin{frontmatter}

\title{Energetic contributions to deformation twinning in magnesium}

\author[BU,UC]{E. Kapan}
\ead{enver.kapan@boun.edu.tr}

\author[BU]{S. Alkan}
\ead{sertan.alkan@boun.edu.tr}

\author[BU]{C. C. Ayd{\i}ner}
\ead{can.aydiner@boun.edu.tr}

\author[UC]{J. K. Mason\texorpdfstring{\corref{cor1}}{}}
\ead{jkmason@ucdavis.edu}
\cortext[cor1]{Corresponding author}

\affiliation[BU]{organization={Mechanical Engineering Department, Bo\u{g}azi\c{c}i University},
            city={Bebek},
            state={Istanbul},
            postcode={34342},
            country={TR}}

\affiliation[UC]{organization={Department of Materials Science and Engineering, University of California, Davis},
            addressline={1 Shields Ave.},
            city={Davis},
            state={CA},
            postcode={95616},
            country={USA}}

\begin{abstract}
Modeling deformation twin nucleation in magnesium has proven to be a challenging task.
In particular, the absence of a heterogeneous twin nucleation model which provides accurate energetic descriptions for twin-related structures belies a need to more deeply understand twin energetics.
To address this problem, molecular dynamics simulations are performed to follow the energetic evolution of $\{10\overline{1}2\}$ tension twin embryos nucleating from an asymmetrically-tilted grain boundary.
The line, surface and volumetric terms associated with twin nucleation are identified.
A micromechanical model is proposed where the stress field around the twin nucleus is estimated using the Eshelby formalism, and the contributions of the various twin-related structures to the total energy of the twin are evaluated. 
The reduction in the grain boundary energy arising from the change in character of the prior grain boundary is found to be able to offset the energy costs of the other interfaces.
The defect structures bounding the stacking faults that form inside the twin are also found to possibly have significant energetic contributions.
These results suggest that both of these effects could be critical considerations when predicting twin nucleation sites in magnesium.
\end{abstract}


\end{frontmatter}

\section{Introduction}
\label{sec:intro}

There continues to be increasing interest in using Mg as a lightweight structural material due to its high strength-to-density ratio and other desirable properties \cite{JOMAgnew2004}.
Mg and Mg alloys are hexagonal close packed (hcp) and accommodate plastic strains by a combination of deformation twinning and slip at room temperature \cite{Agnew2005,ArulKumar2017,Nie2020,Koike2005,WuNature2015}.
Twinning involves a localized finite strain accompanied by a reorientation of the parent crystal, and twinning-dominated plastic deformation causes tension-compression asymmetry and plastic anisotropy at the macroscale due to the unipolar nature of twinning \cite{ArulKumar2017}.
Sample-scale shear banding has been observed \cite{Aydiner2014,Kapan2017,Shafaghi2020} in Mg AZ31 samples with rolling textures loaded in compression perpendicular to the plate normal,
triggering the $\{10\overline{1}2\}$ tension twin mechanism \cite{Nie2020}.
Deformation heterogeneity patterns observed in wrought Mg alloys are also observed to be a function of loading sense \cite{WangRaabe2014, Ucel2019, Orozco-Caballero2017}.
These aspects of the twin nucleation process need to be understood more deeply before rigorous models of Mg plasticity can be developed.

Heterogeneous nucleation is preferred for twin embryos in crystalline Mg due to pronounced energy barrier effects \cite{Yoo1991,Christian1995,Capolungo2008}, with grain boundaries being preferable nucleation sites in polycrystals \cite{WangMRLett2013,BeyerleinJMPS2011}.
Twin nucleation at a grain boundary entails formation of new interfaces (twin boundaries) emanating from the pre-existing grain boundary and is assisted by a set of coordinated dislocation reactions.
The complexity of this process makes predicting possible twin nucleation sites a formidable task and requires accurate energetic descriptions for the twin-related structures.
The purpose of this work is to explore the various energetic contributions to heterogeneous twin nucleation in Mg, laying the groundwork for models to eventually predict heterogeneous twin nucleation sites.

Crystal plasticity (CP) approaches have been popular to model deformation twinning at the continuum scale for more than two decades \cite{PaudelReview2021,Paudel2020}.
Phenomenological formulations usually introduce a twinned volume fraction of material within a pseudo-slip formulation \cite{Roters2010}.
While these approaches are reasonably successful at reproducing the far-field twinning response, they are of limited utility when predicting the nucleation site or the twin embryo morphology \cite{PaudelReview2021,Paudel2020}.
The alternative is to explicitly resolve individual twins as in recently-developed schemes where twin nucleation is modeled as a defect-dissociation reaction, \cite{ChengGhosh2017,ChengGhosh2018} sometimes within phase-field simulations \cite{Kondo2014,Liu2018,Liu2021}.
However, these approaches do not yet consistently incorporate the twin/grain boundary energetics which may play a significant role in heterogeneous twin nucleation.

The possibility of there being other significant energetic contributions that are not included in existing models indicates that there is a need for more detailed atomistic information about the twinning event.
This encouraged the recent characterization of the defect content of twin-parent grain interfaces (i.e., basal-prismatic/prismatic-basal (BP/PB) boundaries and coherent twin boundaries (CTB)) by transmission electron microscopy (TEM) and the subsequent development of atomic-scale models for twin nucleation and growth mechanisms \cite{WangLiu2013,Barrett2014,GongMatResLett2017,Sun2014,Jiang2022}.
These models suggest that the glide of twinning dislocations along a CTB drives the growth of a twin, while the CTB-BP junctions serve as dislocation nucleation and annihilation sites \cite{Ostapovets2014}.
MD calculations by El Kadiri et.\ al.\ \cite{ElKadiri2015} explain the profuse activity of the $\{ 10\overline{1}2\}$ tension twinning mode in Mg by the ability of the twin boundary to absorb basal dislocations without a loss of mobility \cite{ElKadiri2015}.
Liu et.\ al.\ \cite{Liu2016}, using a combination of TEM experiments and MD simulations, characterized CTB and BP/PB facets and showed that the interaction of dislocations with these boundaries plays an important role in twinning energetics;
this is reinforced by subsequent work \cite{WangScience2020, GongMatResLett2017,GongActa2018,GongActa2021,Spearot2020}.

While investigations into the atomic structure of twin interfaces generally focus on pre-existing twin domains, this study is instead concerned with heterogeneous twin nucleation.
As a starting point, electron backscatter diffraction indicates that the misorientation between neighboring grains as well as the grain boundary length are significant factors affecting the statistics of twin nucleation and propagation \cite{BeyerleinJMPS2011,BeyerleinPhilMag2010,Khosravani2015}.
Wang et.\ al.\ \cite{WangScripta2010} and Beyerlein et.\ al.\ \cite{BeyerleinJMPS2011} explored the relationship between GB misorientation and twinning using MD simulations of twins nucleating from a set of symmetrically-tilted grain boundaries (STGBs).
The simulations initiated twin nucleation reactions by introducing basal dislocation pile-ups at the STGBs \cite{WangScripta2010,BeyerleinJMPS2011,WangIntJPlas2013,WangMRLett2013}, and informed a stochastic twin nucleation model where the likelihood of twin nucleation increases with decreasing misorientation angle \cite{BeyerleinJMPS2011}.
Barrett et.\ al.\ \cite{Barrett2014,BarrettActa2014} subsequently investigated the twin nucleation mechanism on a pre-existing BP boundary and identified the relevance of asymmetrically-tilted grain boundaries (ATGB), defect transformation from CTB to BP facets, and the mobility of BP facets to twin nucleation.
Giri et.\ al.\ \cite{Giri2020} used the nudged elastic band method to measure the energy barrier of twin embryo formation on a bi-crystal STGB with a specific misorientation and macroscopic stress condition.
Though all of these studies consider the energetics of twin nucleation, they initiate twin nucleation by intentionally imposing defects or assuming specific reaction pathways.
Alternatively, simulating twin nucleation in a polycrystalline microstructure with MD has been done in a limited number of studies before \cite{Kim2010,Miyazawa2015,Agarwal2019}, but a deep understanding of twin nucleation in these conditions has yet to be established.
Additional studies of spontaneous twin nucleation driven by external loading, without prior assumptions about the nature of the mechanism, could therefore be valuable.

\begin{figure}[ht]
    \centering
    \includegraphics{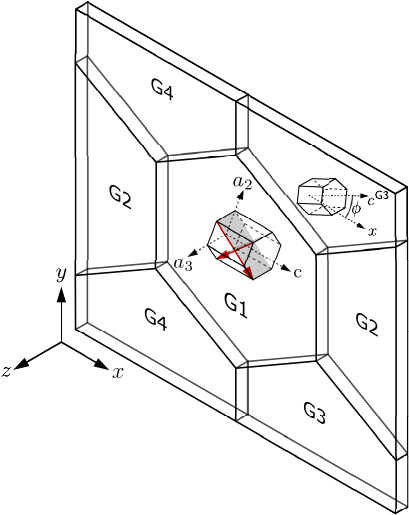}
    \caption{Schematic of the periodic quasi-2D simulation cell with honeycomb grains.
    The $xyz$ coordinate axes are referred to as the global coordinate system in the text.
    The $[\overline{1}\overline{1}20]$ crystal axes of all grains are aligned with the global $z$-axis.
    Grains 2, 3 and 4 are tilted with respect to Grain 1 by a rotation of $\phi$ about the global $z$-axis, i.e., $\phi$ is the angle between the grain's $c$-axis and the global $x$-axis.
    The $(\overline{1}102)$ and $(1\overline{1}02)$ planes associated with the activated $\{10\overline{1}2\}$ tension twin system are highlighted in the central unit cell, with the respective twinning directions indicated by red arrows.}
    \label{fig:simbox}
\end{figure}

This study uses an MD simulation described in Sec.\ \ref{sec:Methods} to follow the energetic evolution of twin embryos nucleating from ATGBs in a honeycomb polycrystalline structure (Fig.\ \ref{fig:simbox}).
Section \ref{sec:results} provides an overview of the twinning events observed during the MD simulation, identifies the twin-related structures, and measures the overall energetic contribution of the twins to the simulation's potential energy.
The relative energetic contributions of line, surface and volumetric terms to the twin nucleation process are evaluated in Sec.\ \ref{sec:analysis}, and the stress field around the twin nucleus is estimated using the Eshelby formalism.

A micromechanical model for the twin formation energy is developed in the same section, with the energy contribution arising from the change in character of the prior GB found to be significant.
The $\mathrm{I_1}$ stacking faults that form inside the twin and the associated defect structures that bound them are also found to possibly make significant energy contributions, amplifying the conclusions of recent experimental \cite{Wang2019, Zhou2021} and numerical studies \cite{Yue2022} that emphasized the role of $\mathrm{I_1}$ stacking faults inside tension twins.
These results suggest that both of these contributions should be included in any comprehensive energy-based twin nucleation model going forward.

\section{Methods}
\label{sec:Methods}

The MD simulations used the LAMMPS code \cite{LAMMPS} with a modified embedded-atom method interatomic potential for Mg \cite{WuCurtinPotent2015}.
The potential was specifically designed to reproduce defect and interface structures and has been widely employed to simulate plastic deformation in pure Mg and Mg-based alloys.

A honeycomb polycrystal with four columnar grains was generated in a simulation box with periodic boundary conditions in all directions (shown in Fig.\ \ref{fig:simbox}) using the Voronoi tessellation algorithm implemented by the Atomsk software \cite{ATOMSK} .
The $z$-axis of the simulation box coincided with the $[\overline{1}\overline{1}20]$ directions of the grains, and together they constituted a $[11\overline{2}0]$-textured system.
The central grain (indicated by G1) was oriented to align its $c$-axis with the global $x$-axis.
Grains 2, 3 and 4 (indicated by G2, G3 and G4) were misoriented with respect to G1 by $15^{\circ}$, $25^{\circ}$ and $48^{\circ}$ rotations about the $z$-axis, with examples of the global and grain coordinate systems and the tilt angle $\phi$ sketched in Fig.\ \ref{fig:simbox}.
Given this geometry, a compressive load along the global $y$-axis would both impose a tensile strain along the $c$-axis of G1 and result in minimal dislocation activity on the basal plane for which the Schmid factor is zero.
Such conditions strongly favor $\{10\overline{1}2\}$ tension twinning in G1, with more dislocation activity expected in the misoriented neighboring grains.
The simulation cell geometry and periodic boundary conditions effectively constrained all dislocation lines to be aligned along the global $z$-axis and the planes of shear of any twins to be parallel with the $xy$ plane.

After initializing the system, the potential energy was minimized using a conjugate gradient method.
The initial minimization was followed by an equilibration step at 0 MPa and 300 K for 50 ps using a Nose-Hoover style isothermal-isobaric (NPT) thermostat \cite{Shinoda2004} to relax any remaining high energy GB structures.
Each component of the average stress tensor was below 6 MPa after the equilibration step, well below the reported critical stresses for similar structures \cite{Kim2010}.
The structure was then compressed in the $y$-direction at a constant engineering strain rate of $-2\times10^{8}\ \mathrm{s}^{-1}$ for 160 ps.
41 load points (LPs) were taken at 1 ps intervals between 104 ps and 145 ps after the start of compression (corresponding to far-field strains $\varepsilon_{yy}^0$ of $-0.0208$ and $-0.0286$) and will be referred to as LP0 to LP40 below.

\begin{figure*}[ht]
    \centering
    \includegraphics[width=0.75\textwidth]{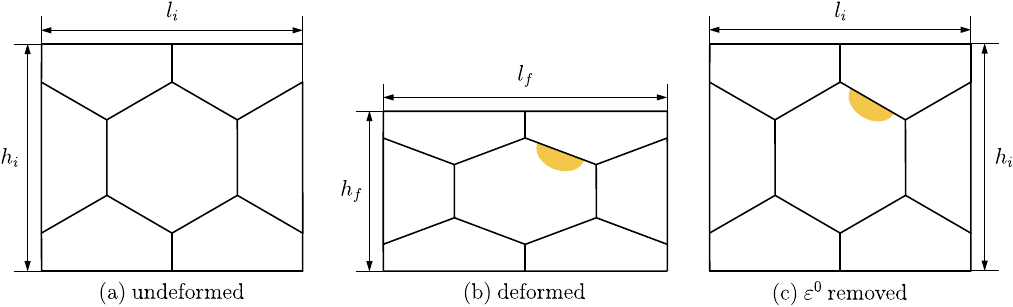}
    \caption{A sketch of (a) the simulation cell after initial equilibration, (b) the deformed simulation cell with a twin in G1 (the deformation is exaggerated), (c) the simulation cell after the removal of far-field strain.
    The initial dimensions $l_i$, $h_i$, $t_i$ are 692 $\text{\AA}$, 600 $\text{\AA}$ and 25.6 $\text{\AA}$ respectively.}
    \label{fig:steps-schematics}
\end{figure*}

After completing the simulation, each LP was instantaneously quenched to 0 K by setting the atomic momenta to zero and minimizing the potential energy.
The far-field strains on each LP were then removed by restoring the simulation box dimensions to those of the undeformed state and mapping the atomic coordinates with the corresponding affine transformation.
This was followed by a final energy minimization to relax any local elastic strains introduced by the affine transformation.
The second minimization did not noticeably affect the shape or extent of the twin nucleus, despite the untwinned structure having a lower potential energy in the undeformed state.
The resulting atomic data (potential energies, atomic coordinates, etc.) were stored for each of the selected LPs for subsequent analysis.
A graphical description of this loading/remapping sequence is given in Fig.\ \ref{fig:steps-schematics}.

The purpose of the minimizations and affine transformation was to remove any thermal fluctuations and far-field elastic strains which would complicate the calculation of the potential energy of the developing twin nucleus.
More specifically, the intention was to examine the potential energy of the twin nucleus in a fixed reference state where the far-field strain vanished, conceptually resembling Eshelby's thought experiment of embedding an inhomogeneity into an infinite medium.

Post-processing and visualization of the atomic data was done using the Open Visualization Tool (OVITO) \cite{Stukowski2010} complemented with custom Python scripts.
The OVITO implementations of Polyhedral Template Matching (PTM) and the Dislocation Extraction Algorithm (DXA) were used to identify crystal structures, interfaces and dislocations.
A grain segmentation scheme based on local (per atom) orientations returned by PTM was implemented to automate the identification of atoms belonging to the twin.

\section{Results}
\label{sec:results}

\begin{figure*}
    \centering
    \includegraphics[width=\textwidth]{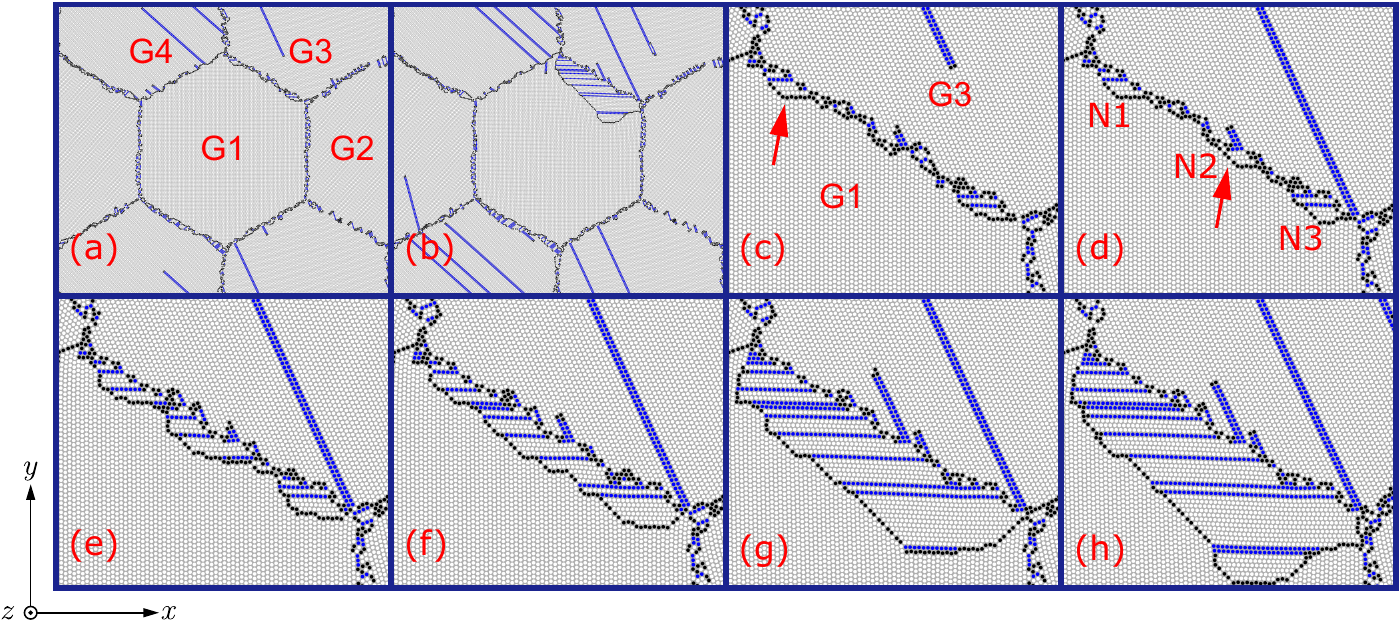}
    \caption{Snapshots of the simulation box and twin nuclei at representative states of the MD simulation after the removal of the far-field strain, with the reported strain values given in reference to the corresponding load step.
    (a,b) Snapshots of the simulation box at LP0 ($\varepsilon_{yy}^0 = -0.0208$) and LP40 ($\varepsilon_{yy}^0 = -0.0286$).
    (c-h) Snapshots of the twinned G1-G3 boundary at LP0, LP9, LP19, LP24, LP32 and LP40 ($t = 104$, $113$, $123$, $128$, $136$ and $144$ ps respectively).
    N1, N2 and N3 indicate the three distinct twin nuclei on the G1-G3 boundary at $113$ ps.
    White and blue indicate atoms in hcp and fcc environments, whereas black denotes a disordered structure on an interface.}
    \label{fig:summary_fig}
\end{figure*}

The stages of the most significant twinning event that occurred during the MD simulation are shown in Fig.\ \ref{fig:summary_fig}, with Figs.\ \ref{fig:summary_fig}a and b showing the entire simulation cell at the beginning and end of the event.
The LP0 structure in Fig.\ \ref{fig:summary_fig}c contained two twin embryos on the G1-G3 boundary, with one (upper left, indicated by a red arrow) nucleating around LP0 and the other (bottom right) nucleating during the initial equilibration step before any external straining.
By LP9 in Fig.\ \ref{fig:summary_fig}d, a third twin nucleus indicated by a red arrow appeared between the two preexisting nuclei.
Figure \ref{fig:summary_fig}e shows the coalescence of N1 and N2 at LP19, and all of the nuclei coalesced into a single large twin by LP24 in Fig.\ \ref{fig:summary_fig}f.
The twin had extended along the entire prior GB and continued to expand into the interior of G1 by LP32 in Fig.\ \ref{fig:summary_fig}g.
The twin at the final loading point LP40 in Fig.\ \ref{fig:summary_fig}h developed new PB and CTB facets on the lower right, and overcame the G1-G2-G3 triple junction barrier to extend along the G1-G2 boundary.
The final twin is clearly well-developed and stable, and would continue to expand into the interior of the grain had the simulation continued.
The purpose here is to investigate the initial stages of twin formation though, so the later stages of intragranular twin growth will not be considered.
Instead, the analysis that follows will be limited to the twin formation stages on the G1-G3 grain boundary visible in Fig.\ \ref{fig:summary_fig}.

\subsection{Energetic Analysis}
\label{subsec:energetic_analysis}

\begin{figure}[ht]
    \centering
    \includegraphics{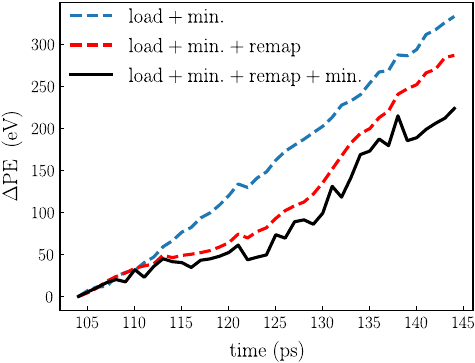}
    \caption{The change in potential energy of the simulation cell relative to values at LP0 ($t = 104$ ps) (i) after the first energy minimization, (ii) after removal of the far-field strain $\varepsilon^0$, (iii) after the second energy minimization.}
    \label{fig:pes_vs_time}
\end{figure}

The change in potential energy $\Delta PE$ of the load points with respect to LP0 are calculated and plotted in Fig.\ \ref{fig:pes_vs_time}.
The potential energy generally increases from LP0 to LP40 with the accumulation of plastic strain and various defect structures.
The three curves in the figure correspond to the potential energies of the load points after each of the three post-processing operations described in Sec.\ \ref{sec:Methods}.
The dashed blue curve shows the change in potential energy after the first energy minimization, i.e., after instantaneously quenching the structure to 0 K.
The dashed red curve shows the decrease in the potential energy after removing the far-field strain by means of the affine transformation.
The solid black curve is obtained after the second energy minimization relaxes the local elastic strains introduced by remapping the atomic coordinates.
That is, the solid black curve should reflect only the effect of introducing the twin and other defects into the system's reference configuration.
The advantage of this construction is that the far-field stress that drives the nucleation of the twin is removed, allowing the energetic model for twin nucleation to be simplified by dropping one of the terms.
This also allows the remaining elastic strain energy to more readily be approximated by Eshelby's solution to the generalized inhomogeneity problem \cite{Eshelby1957}.

\begin{figure}
    \centering
    \includegraphics{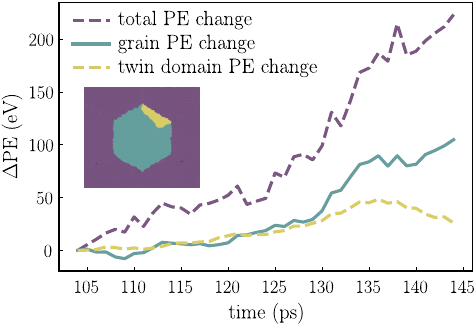}
    \caption{The change in potential energy relative to the values at LP0 ($t = 104$ ps) of (i) the entire simulation cell, (ii) G1, and (iii) the region occupied by the twin at $140$ ps.}
    \label{fig:total_vs_twin_pe}
\end{figure}

While the energetic analysis of the developing twin nucleus could be performed using the black curve in Fig.\ \ref{fig:pes_vs_time}, accounting for the energetic contributions of the various defect structures that develop in G2, G3 and G4 would introduce complications.
For example, the $1/3\langle\overline{1}100\rangle$ Shockley partials in G3 and G4 in Fig.\ \ref{fig:summary_fig} contribute dislocation core and elastic strain energies, and the trailing $\mathrm{I}_{2}$-type stacking faults (visible in blue in Fig.\ \ref{fig:summary_fig}a-b) have a surface energy that would need to be removed from the overall potential energy change.
Restricting the analysis to only the region occupied by the twin at LP40 would instead introduce complications from the compatibility strain field surrounding the twin gradually extending outside of the region as the nucleus develops.
The significance of this effect can be seen in Fig.\ \ref{fig:total_vs_twin_pe} where the solid green curve shows the potential energy change of G1 (the union of green and yellow regions in the inset) and the dashed yellow curve shows only that of the region occupied by the twin at LP40 (the yellow region in the inset).
The potential energy changes are similar when the twin nuclei are small compared to the yellow region, but the yellow curve falls below the green curve as the twin size becomes comparable to that of the yellow region and more of the energy stored in the strain field external to the twin is excluded.
The simplest option seems to be to consider the potential energy change of G1, with a comparison of Figs.\ \ref{fig:summary_fig}a and \ref{fig:summary_fig}b showing that G1 has limited defect activity elsewhere in the grain during the relevant time interval (likely owing to the vanishing of the Schmid factor for basal slip).
The potential energy change inside G1 is therefore ascribed to the developing twin nucleus, and the green curve in Fig.\ \ref{fig:total_vs_twin_pe} is selected as the basis for the following analysis.
It should be noted though that the green curve is not the overall potential energy of the twin nucleus since the energy contribution of the compatibility strains that develop in the neighboring G3 is neglected.
The neglected energy will be accounted for by an approximation introduced in Sec.\ \ref{subsec:eshelby} below.

\subsection{\texorpdfstring{$\{10\overline{1}2\}$}{\{10-12\}} Twin Related Structures}
\label{subsec:twin_related_structures}

\begin{figure}[ht]
    \centering
    \includegraphics{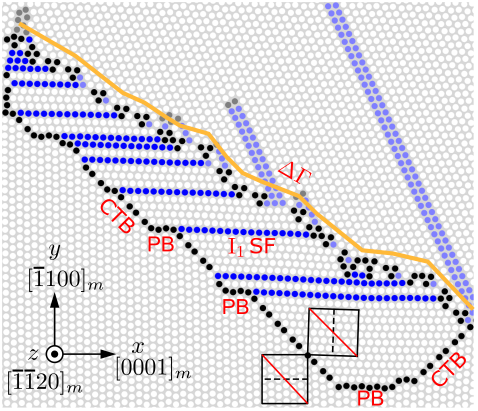}
    \caption{Interfacial energy contributions associated with a $\{10\overline{1}2\}$ twin include those for coherent twin boundaries (CTBs), basal-prismatic boundaries (PBs), $\mathrm{I_1}$ stacking faults ($\mathrm{I_1}$ SFs) inside the twin, and the energy change $\Delta \Gamma$ of the prior GB.
    White and blue indicate atoms in hcp and fcc environments, whereas black denotes a disordered structure on an interface.}
    \label{fig:twin_str}
\end{figure}

The general structure of the twin observed in the MD simulation is shown in Fig.\ \ref{fig:twin_str} with the characteristic structures at LP28 highlighted and labeled.
The cross-section of a $\{10\overline{1}2\}$ twin viewed along the $[\overline{1}\overline{1}20]$ axis usually has a stepped boundary consisting of CTB and PB/BP facets.
A CTB facet is one where the plane of the boundary coincides with the twin plane (indicated in the following by $K_1$), and is the $\{10\overline{1}2\}$ plane for a $\{10\overline{1}2\}$ twin.
The orientation of the $K_{1}$ plane is indicated with red lines inside the representative hcp unit cells sketched for the matrix and twin in Fig.\ \ref{fig:twin_str}.
There is a negligible misalignment of $\about 2^\circ$ between the actual CTB and the expected CTB orientations, likely due to local atomic strains.
A PB/BP facet is one where the prismatic/basal plane of the matrix adjoins the basal/prismatic plane of the twin.
In Fig.\ \ref{fig:twin_str}, the facets labeled PB are those where the basal planes of the twin coincide with the prismatic planes of the matrix.
Both the basal-prismatic and prismatic-basal boundaries will be indicated by PB in the following since the two have essentially identical characters.
Recent studies suggest that the growth of a twin into the interior of the matrix is facilitated by the formation of a terrace (a disconnection pair) on the CTB \cite{Ostapovets2014} and subsequent glide of the disconnections along the CTB \cite{Yue2022}, thereby involving an interplay of CTB and PB boundaries.

\begin{figure}[ht]
    \centering
    \includegraphics[width=0.9\columnwidth]{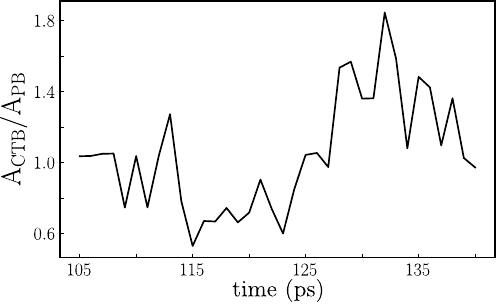}
    \caption{The ratio of total CTB facet area to total PB facet area on the boundary of the twin nucleus for all load points LP0 to LP36.}
    \label{fig:ctb_vs_ictb}
\end{figure}

As a brief aside, the historical assumption has been that a twin embryo is initially bounded predominantly or even entirely by CTB facets 
\cite{Pond2016}.
However, recent HREM studies \cite{Jiang2022} have established that a twin embryo boundary can have a significant proportion of BP/PB facets.
The ratio of total CTB facet area to total PB facet area in the MD simulation is plotted in Fig.\ \ref{fig:ctb_vs_ictb} for all load points LP0 to LP36, and is consistent with roughly half of the twin boundary being PB facets. 

Two other structures are labeled in Fig.\ \ref{fig:twin_str} in addition to the twin-matrix boundaries.
The first are $I_{1}$ stacking faults (SFs) that change the stacking order of the basal planes to $AB\underline{A}CA$ (or equivalently $BA\underline{B}CB$);
recent high resolution electron microscopy (HREM) studies have verified the presence of such SFs inside $\{10\overline{1}2\}$ twins \cite{Wang2019,Zhao2019,Zhou2021}.
The SFs usually extend from the twin boundary on one side to the GB on the other, with the intersection with the twin boundary frequently occurring at a junction between CTB and PB facets.
The second structure is the traced boundary between the twin and the neighboring G3 that is labeled with a $\Delta \Gamma$.
The GB initially between G1 and G3 effectively decomposed with the emission of the twin boundary, leaving behind a GB of changed character.
Similar decomposition reactions can be seen to have been initiated on several other GBs (e.g., the boundaries between G1 and G2) in Fig.\ \ref{fig:summary_fig}a.

\begin{figure}[ht]
    \centering
    \includegraphics[]{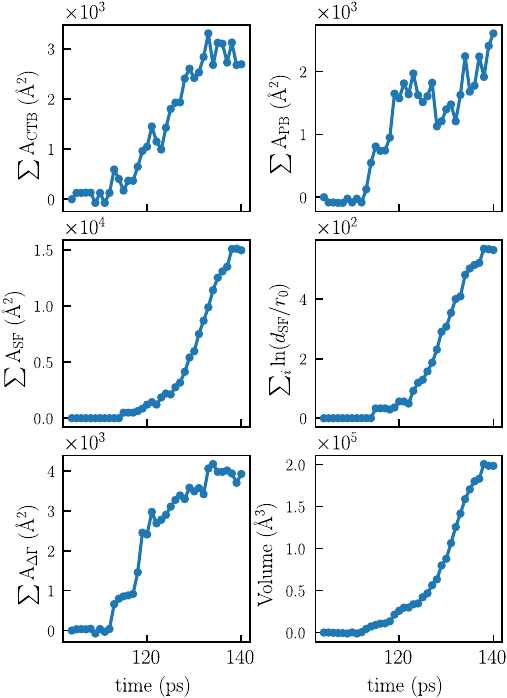}
    \caption{Magnitudes of the various $\{10\overline{1}2\}$ twin related structures.
    In addition to changes in the CTB, PB, SF and $\Delta\Gamma$ areas, changes to the twin nucleus volume and the sum of the natural logarithms of stacking fault lengths (normalized by the twice the lattice constant of Mg) are reported.}
    \label{fig:md_pars_evol}
\end{figure}

The magnitudes of the various twin related structures are quantified by direct measurement and reported in Fig.\ \ref{fig:md_pars_evol}.
Lengths are calculated as the sum of the lengths of line segments connecting adjacent atomic columns, and areas are calculated by multiplying a length by the depth of the simulation cell.
Volumes are calculated as the sum of the volumes of the Voronoi cells around the relevant atoms.
These quantities are used extensively for the analysis in Sec.\ \ref{sec:analysis} below.

\section{Analysis and Discussion}
\label{sec:analysis}

The total potential energy of the twin nucleus is likely composed of contributions from the interfaces, the defect substructure, and a strain energy arising from the transformation strain.
Although a precise partitioning of the twin potential energy between these contributions is almost intractable due to the number of unknowns and the difficulty of mapping concepts like an interface or a dislocation core to a specific set of atoms, a micromechanical model for the potential energy can nevertheless be developed by applying certain simplifications.
As a starting point, a model is proposed for the change in the potential energy of G1 that includes terms for the twin boundary facets, the stacking faults, the prior grain boundary, and a term that is proportional to the twin volume:
\begin{align}
\label{eq:model_1}
    \Delta E_\mathrm{G1} &= \gamma_\mathrm{CTB} A_\mathrm{CTB} + \gamma_\mathrm{PB} A_\mathrm{PB} + \gamma_\mathrm{SF} A_\mathrm{SF} \nonumber \\
    &\quad + \gamma_{\Delta \Gamma} A_{\Delta \Gamma} + C_\mathrm{V} V.
\end{align}
The CTB, PB and SF contributions are modeled as the products of the equilibrium excess surface energies (148, 164 and 11 $\mathrm{mJ/m^2}$ respectively \cite{WuCurtinPotent2015, WuActa2016, DangActa2020}) and the corresponding interfacial areas.

The $\Delta \Gamma$ term accounts for the change in the surface energy of the GB initially between G1 and G3 with the emission of the twin boundary.
$\gamma_{\Delta \Gamma}$ will be estimated by a linear regression analysis in Sec.\ \ref{subsec:regression}, with the resulting value verified by a molecular statics calculation in Sec.\ \ref{subsec:GB_decomp}.

The volumetric term $C_\mathrm{V} V$ will initially be assumed in Secs.\ \ref{subsec:PCA} and \ref{subsec:eshelby} to be proportional to the elastic strain energy introduced by applying a transformation strain to an elliptic cylinder inhomogeneity in an infinite elastic medium.

This model will eventually be found to be unable to account for the potential energy of the twin nucleus, resulting in a revision of the model in Sec.\ \ref{subsubsec:updating_model} to include the possible elastic interactions of disconnections in the twin boundary and the prior GB (Fig.\ \ref{fig:md_pars_evol} shows that any potential energy stored in these interactions would be nearly proportional to the twin volume).
The final analysis strongly suggests that the energy change of the prior GB, and perhaps the elastic interactions of the disconnections, contribute significantly to the potential energy of the nucleating twin.

\subsection{Regression Analysis}
\label{subsec:regression}

The CTB, PB, and SF terms in Eq.\ \ref{eq:model_1} are determined by the equilibrium excess surface energies reported in the literature and the interfacial areas in Fig.\ \ref{fig:md_pars_evol}.
This leaves $\gamma_{\Delta \Gamma}$ and $C_\mathrm{V}$ as the only unknown quantities in Eq.\ \ref{eq:model_1}, and since they appear linearly in the equation, they can be estimated by linear least squares using the potential energies of G1 at each of the $N$ loading points.
Explicitly, the linear regression problem takes the following form:
\begin{equation}
\label{eq:regression1}
    \begin{bmatrix}
    A_{\Delta \Gamma}^{\mathrm{LP}0} & V^{\mathrm{LP}0}\\
    \vdots & \vdots \\
    A_{\Delta \Gamma}^{\mathrm{LP}N} & V^{\mathrm{LP}N}\\
    \end{bmatrix}
    \begin{bmatrix}
    \gamma_{\Delta \Gamma} \\
    C_\mathrm{V}
    \end{bmatrix}
    =
    \begin{bmatrix}
    \Delta E_\mathrm{G1}^0 - \sum_{i}\gamma_{i}^0 A_{i}^0\\
    \vdots \\ 
    \Delta E_\mathrm{G1}^N - \sum_{i}\gamma_{i}^N A_{i}^N
    \end{bmatrix}
\end{equation}
where the sum on the right hand side includes the CTB, PB and SF interfaces.
Solving this gives $-0.140\pm 0.011 \ \mathrm{J/m^2}$ and $52.6\pm 3.31\ \mathrm{MJ/m^3}$ for the parameters $\gamma_{\Delta \Gamma}$ and $C_\mathrm{V}$, respectively.

\begin{figure}
    \centering
    \includegraphics{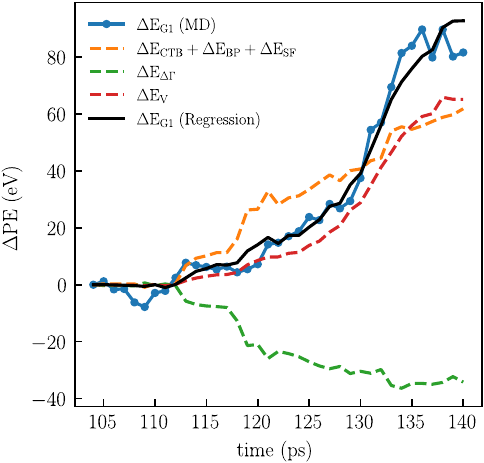}
    \caption{Energetic contributions of the various terms involved in the regression analysis in Eq.\ \ref{eq:regression1} are plotted with dashed lines.
    The change in the G1 potential energy during the MD simulation (blue curve) is used as a measure of the change in the twin potential energy.
    The black curve is the sum of the dashed curves and is the least squares fit to the blue curve.}
    \label{fig:regression1}
\end{figure}

The results of the regression analysis are plotted in Fig.\ \ref{fig:regression1} alongside the MD measurements for comparison.
It is significant that the $\Delta E_{\Delta \Gamma}$ curve shows a significant decrease in the energy of the G1-G3 GB as the twin propagates along the GB.
Note also that $C_\mathrm{V}$ here includes the energetic contributions from all possible sources that scale with the twin volume, including but not restricted to the elastic strain energy.
The elastic strain energy will be estimated by the Eshelby formalism in the following sections, allowing an evaluation of whether this is sufficient to explain the $\Delta E_\mathrm{V}$ curve in Fig.\ \ref{fig:regression1}.

\subsection{Principal Component Analysis}
\label{subsec:PCA}

\begin{figure*}
    \centering
    \includegraphics[width=\textwidth]{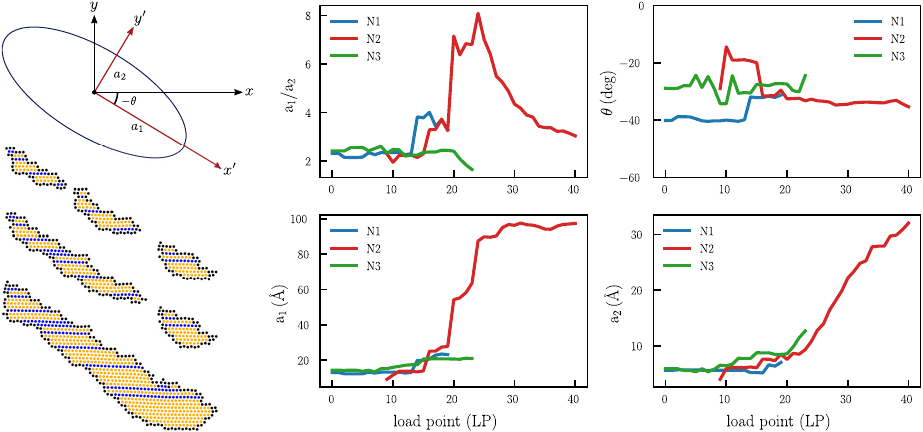}
    \caption{The evolution of the the geometric properties of the ellipses fitted to the twin nuclei.
    All twin nuclei coalesce by LP24 and only the geometric data for the ellipse fitted to N2 is reported for subsequent load points.}
    \label{fig:nuclei_pars}
\end{figure*}

The calculation of the elastic strain energy associated with a twin in the Eshelby inhomogeneity theory requires that the geometric parameters of the twinned regions be known to first obtain the Eshelby tensor.
A twin is modeled as an elliptic cylinder since twins are frequently observed to have elliptic morphologies during the initial twinning stages \cite{Barrett2014b}.
At each load point, the atoms that belong to each of the twinned regions are identified, and the dimensions of the elliptic cylinder that contains the associated point cloud are found using principal component analysis.
Specifically, the atomic coordinates matrix is defined as
\begin{equation}
    \mathbf{A} = 
    \begin{bmatrix}
    x_{1}&y_{1} \\
    x_{2}&y_{2} \\
    \vdots&\vdots
    \end{bmatrix}.
\end{equation}
The directions of the semi-major and semi-minor axes are taken to be the directions of the first and second eigenvectors of $\mathbf{A}^T \mathbf{A}$.
If $a'_1$ and $a'_2$ are the square roots of the corresponding eigenvalues, then the major and minor radii $a_1$ and $a_2$ are defined by the conditions that $a_1 / a_2 = a'_1 / a'_2$ and the volume of the resulting elliptical cylinder is identical to the twin volume.
The orientation of the elliptic cylinder with respect to the simulation cell's coordinate system is expressed as the angle $\theta$ between the semi-major axis and the $x$-axis.
The major and minor radii, the aspect ratios and the orientations of the twins are all reported in Fig.\ \ref{fig:nuclei_pars}.
All three of the twin nuclei coalesce by LP24, leaving only the geometric parameters of the second nucleus in the plots beyond this loading point.
The geometric quantities $a_{1}$, $a_{2}$ and $\theta$ are used below to model the twins as elastic inhomogeneities and estimate the associated elastic strain energies.

\subsection{Elastic Field Model}
\label{subsec:eshelby}

Eshelby showed that the total strain energy $E_\mathrm{inh}$ in an infinite elastic medium due to a transformed inhomogeneity embedded inside the medium is equal to:
\begin{equation}
    E_\mathrm{inh}=-\frac{1}{2}V\sigma_{ij}\varepsilon_{ij}^p
\end{equation}
where $\sigma_{ij}$ is the total stress inside the inhomogeneity, $\varepsilon_{ij}^p$ is the eigenstrain associated with the twin transformation, and the Einstein summation convention is followed.
$E_\mathrm{inh}$ is the total strain energy in the entire medium and can be decomposed into two parts, namely, the elastic strain energies stored in the interior and exterior of the inhomogeneity \cite{BarnettCaiJMPS2018}:
\begin{align}
    E_\mathrm{inh} &= E_\mathrm{int} + E_\mathrm{ext} \nonumber \\
    &= \frac{1}{2} V\sigma_{ij} (\varepsilon_{ij} - \varepsilon_{ij}^p) - \frac{1}{2} V\sigma_{ij} \varepsilon_{ij}
    \label{eq:inh_e_decomp}
\end{align}
where $\varepsilon_{ij}$ and $\varepsilon_{ij} - \varepsilon_{ij}^p$ are the total and elastic strains in the inhomogeneity, respectively.
Eshelby's equivalent inclusion theory \cite{Eshelby1957,mura2013micromechanics} allows the total stress and strain to be found as $\sigma_{ij} = \left(C_{ijkl} S_{klmn} - C_{ijmn}\right) \varepsilon_{mn}^{**}$ and $\varepsilon_{ij} = S_{ijkl}\varepsilon_{kl}^{**}$ where $S_{ijkl}$ is the Eshelby tensor and $\varepsilon_{kl}^{**}$ is the equivalent eigenstrain tensor in the equivalent inclusion problem.
$\varepsilon_{ij}^{**}$ is related to $\varepsilon_{ij}^p$ by:
\begin{equation}
[(C_{ijkl}^* - C_{ijkl})S_{klmn} + C_{ijmn}] \varepsilon_{mn}^{**} = C_{ijkl}^*\varepsilon_{kl}^p
\end{equation} 
where $C_{ijkl}^*$ and $C_{ijkl}$ are the stiffness tensors of the twin and the surrounding medium, respectively.

The eigenstrain tensor $\boldsymbol{\varepsilon}^p$ associated with the $\{10\overline{1}2\}$ Mg twin in the simulation cell coordinate system is:
\begin{equation}\label{eq:eps_p}
    \boldsymbol{\varepsilon}^p
    =
    \begin{bmatrix}
        0.065 & 0 & 0 \\
        0 &-0.065 & 0 \\
        0 & 0 & 0
    \end{bmatrix}
\end{equation}
The medium is modeled as having the same isotropic elastic response as polycrystalline Mg \cite{CapolungoIsotropic2010, TromansIsotropic2011}; this is not regarded as a severe approximation since the medium is composed of magnesium single crystals with multiple orientations, each of which is nearly elastically isotropic.
The equivalent inclusion (i.e., the twin) is instead modeled as being transversely isotropic in the basal plane, consistent with hcp crystal symmetry.
The elastic properties for both the medium and the twins are reported in Table \ref{tab:elastic_props}.
The components of the Eshelby tensor for an elliptic cylinder are functions of the major and minor radii $a_1$ and $a_2$ and the elastic constants of the medium, and these functions have been tabulated by Mura \cite{mura2013micromechanics}.

\begin{table}
    \centering
    \begin{tabular}{p{15mm} p{15mm} p{15mm} p{15mm}}
    \toprule
    \multicolumn{2}{c}{Medium (GPa)} & \multicolumn{2}{c}{Twin (GPa)} \\
    \hline
    $E$& 45&  $C_{1111}^{*}$& 63.5\\
    $\nu$& 0.28&  $C_{1122}^{*}$& 25.9\\
    $\mu$& $17.6$&  $C_{1133}^{*}$& 21.7\\
    &  &  $C_{3333}^{*}$& 66.5\\
    &  &  $C_{1313}^{*}$& 18.4\\
    \bottomrule
    \end{tabular}
    \caption{Elastic constants of the medium and the twin in standard orientations.
    The components of the stiffness tensor of the medium can be obtained from the isotropic elastic properties.}
    \label{tab:elastic_props}
\end{table}

\begin{figure}
    \centering
    \includegraphics[width=0.8\columnwidth]{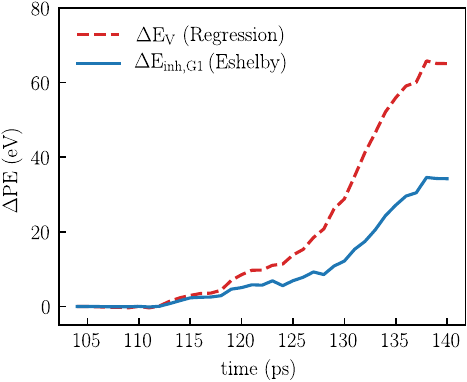}
    \caption{The volumetric part of the twin energy ($\mathrm{\Delta E_\mathrm{V}}$, dashed red curve) as approximated by the regression analysis (Eq.\ \ref{eq:regression1}) is compared with the elastic strain energy predicted $E_{\mathrm{inh,G1}}$ by the Eshelby solution (Eq.\ \ref{eq:inh_e_G1}, solid blue curve).}
    \label{fig:vol_terms_comp_1}
\end{figure}

After solving for the elastic fields associated with the inhomogeneity, $E_\mathrm{inh}$ for the Eshelby solution can be compared to the volumetric energy term obtained from the regression results.
However, the energy change of G1 is used as the objective for the regression analysis in Eq.\ \ref{eq:regression1}, whereas the Eshelby solution accounts for the total elastic strain energy stored in the entire medium.
This makes Eq.\ \ref{eq:inh_e_decomp} not directly comparable to the regression results.
The elastic energy stored in G1 is instead modeled as the sum of the elastic strain energy stored on the interior and half that stored on the exterior of the inhomogeneity:
\begin{align}
    E_\mathrm{inh,G1} &= E_\mathrm{int} + E_\mathrm{ext,G1} \nonumber \\
    &\approx \frac{1}{2} V\sigma_{ij} (\varepsilon_{ij} - \varepsilon_{ij}^p) - \frac{1}{4} V\sigma_{ij} \varepsilon_{ij}
    \label{eq:inh_e_G1}
\end{align}
The prediction of Eq.\ \ref{eq:inh_e_G1} is compared to the volumetric energy term obtained from the regression analysis in Fig.\ \ref{fig:vol_terms_comp_1}, where the volumetric term obtained from the regression analysis is seen to be significantly higher than that predicted by the Eshelby solution.
The discrepancy does not necessarily suggest a failure of the energy model in Eq.\ \ref{eq:model_1} or of the approximations involved in our application of the Eshelby solution, but rather that there could be an additional energy contribution that scales with the volume of the twin and is distinct from the elastic strain energy in the Eshelby solution.

\subsubsection{Revising the Micromechanical Model}
\label{subsubsec:updating_model}

\begin{figure}[ht]
    \centering
    \includegraphics[]{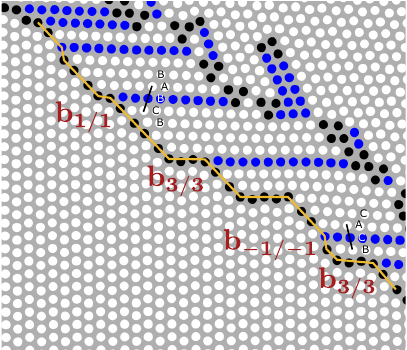}
    \caption{$\mathrm{I_1}$ basal stacking faults inside the $\{10\overline{1}2\}$ twin.
    Each SF is bounded by a defect pair that anchors the SF to the twin boundary on the left and the grain boundary on the right.
    Defects in the TB are associated with short PB or BP steps on the CTB and have disconnection Burgers vectors denoted by $\mathrm{\mathbf{b}_{\pm\mu/\pm\kappa}}$ with either two positive or two negative subscripts.
    These correspond to two different changes in the stacking sequence, with the stacking sequence changes associated with $\mathrm{\mathbf{b}_{1/1}}$ and $\mathrm{\mathbf{b}_{-1/-1}}$ disconnections identified in the figure.}
    \label{fig:sf_closeup}
\end{figure}

\begin{figure}
    \centering
    \includegraphics[width=0.8\columnwidth]{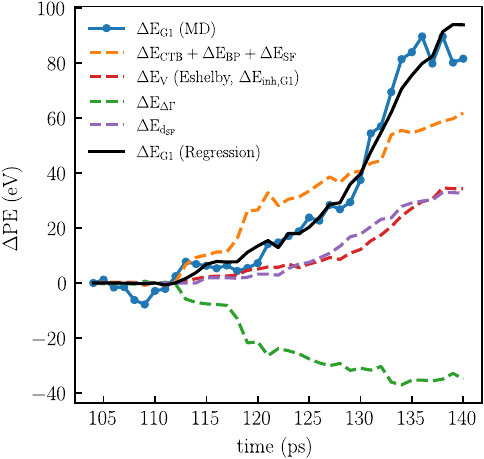}
    \caption{Energetic contributions of the various terms involved in the second regression analysis in Eq.\ \ref{eq:regression2} are plotted with dashed lines.
    The change in the G1 potential energy during the MD simulation (blue curve) is used as a measure of the change in twin potential energy.
    The black curve is the sum of the dashed curves and is the least squares fit to the blue curve.}
    \label{fig:regression2}
\end{figure}

A closer examination of the defect substructure inside the twin in Fig.\ \ref{fig:sf_closeup} reveals that the point where a stacking fault intersects the twin boundary almost invariably coincides with a corner of a BP or PB facet.
Such stacking faults have also been observed experimentally in $\{10\overline{1}2\}$ twins \cite{Wang2019, Zhou2021}, and their possible origins and the character of their bounding defects have been studied in detail \cite{Yue2022, WangScripta2018, HeActa2020}.
The bounding defect in the twin boundary is specifically known to have disconnection character;
the Burgers vectors and the step heights of the disconnections in Fig.\ \ref{fig:sf_closeup} are given in the conventional notation $\mathbf{b}_{\pm \mu/\pm \kappa}$ used in interfacial defect theory \cite{Pond1999}.
The sign of the subscripts $\mu$ and $\kappa$ indicate a step ``up'' or ``down'' with respect to the parent crystal, and their magnitudes signify the height of the step.
The other bounding defect of each stacking fault is embedded in the GB and has a Burgers vector determined by the initial defect content of the GB and the dissociation reaction that produces the associated SF.

The model in Eq.\ \ref{eq:model_1} presumably includes the self-energies of the defects at the ends of the stacking faults in the effective energies of the bounding interfaces, but does not include the interaction energies of those defects.
The conservation of Burgers vector content implies that the defects at either end of a stacking fault have equal and opposite Burgers vectors and hence exert attractive forces per unit length on one another.
By analogy with the work required to separate a full dislocation into a pair of Shockley partials, the work required to separate the defects at either end of the $i$th stacking fault is modeled as being proportional to $\ln(r_i / r_0)$ where $r_i$ is the length of the stacking fault, $r_0 = 2 a$ is a reference length scale, and $a$ is the lattice constant of Mg.
Including this interaction term in the model for the change of the potential energy of G1 gives:
\begin{align}
\label{eq:model_2}
 \Delta E_\mathrm{G1} &= \gamma_\mathrm{CTB} A_\mathrm{CTB} + \gamma_\mathrm{PB} A_\mathrm{PB} + \gamma_\mathrm{SF} A_\mathrm{SF} \nonumber \\
    &\quad + E_\mathrm{inh,G1} + \gamma_{\Delta \Gamma} A_{\Delta \Gamma} + K t \sum_{i} \ln(r_i / r_0)
\end{align}
where $t$ is the thickness of the simulation cell and the prefactor $K$ has units of $\text{eV/\AA}$.
While the SFs inside the twins are bounded by different defect pairs with different Burgers vectors and the prefactors associated with each defect pair should therefore be different, $K$ is an average prefactor that accounts for the total interaction energy between the arrays of defects on the TB and on the GB.

With this modification, the linear regression problem to find the two unknown parameters in Eq.\ \ref{eq:model_2} becomes:
\begin{align}
    &\begin{bmatrix}
        A_{\Delta \Gamma}^{\mathrm{LP}0} & t \sum_{i} \ln\left(\frac{r_{i}^{0}}{r_0}\right)\\
        \vdots & \vdots\\
        A_{\Delta \Gamma}^{\mathrm{LP}N} & t \sum_{i} \ln\left(\frac{r_{i}^{N}}{r_0}\right)\\
    \end{bmatrix}
    \begin{bmatrix}
        \gamma_{\Delta \Gamma} \\
        K
    \end{bmatrix} = \vec{y}
    \label{eq:regression2}
\end{align}
where $\vec{y}$ is the column vector whose $j$th entry is given by $\Delta E_\mathrm{G1}^j - \sum_{i}\gamma_{i}^j A_{i}^j - E_{\mathrm{inh,G1}}^j$.
Solving this gives $-0.142\pm 0.013\ \mathrm{J/m^2}$ and $0.0578\pm 0.0077\ \mathrm{eV}/\text{\AA}$ for the parameters $\gamma_{\Delta \Gamma}$ and $K$, respectively.
The value for $\gamma_{\Delta \Gamma}$ is nearly the same as that obtained in Sec.\ \ref{subsec:regression} using Eq.\ \ref{eq:regression1}, and will be discussed in more detail in Sec.\ \ref{subsec:GB_decomp}.
The results of the regression analysis are plotted in Fig.\ \ \ref{fig:regression2} alongside the MD measurements for comparison.

The estimated value of $K$ should be comparable to the prefactor for a single pair of typical defects that bound an $\mathrm{I_1}$ SF segment inside $\{10\overline{1}2\}$ twins.
The Burgers vectors of the PB disconnections on the twin boundary have varying magnitudes between $0.150a$ and $0.350 a$ \cite{Yue2022}.
The $\langle c+a \rangle$ dislocations embedded in the grain boundary on the other end of the stacking fault have Burgers vectors of length $\about a$.
For the average case, the prefactor of the interaction energy between a dislocations with Burgers vector lengths of $b_1 = 0.250a$ and $b_2 = a$ respectively is roughly:
\begin{equation}
    K \propto \frac{\mu b_1 b_2}{2\pi} = 0.0409\ \mathrm{eV}/\text{\AA}
\end{equation}
which is sufficiently close to the least square estimate for the latter to be reasonable.
The only remaining term in Eq.\ \ref{eq:model_2} that has not been shown to be physically reasonable is the $\gamma_{\Delta \Gamma} A_{\Delta \Gamma}$, and this is considered in detail below.

\subsection{Grain Boundary Decomposition}
\label{subsec:GB_decomp}

\begin{figure}
    \centering
    \includegraphics[width=\linewidth]{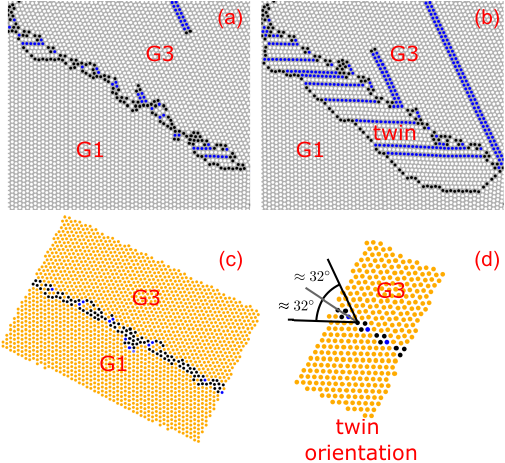}
    \caption{(a) Snapshot of the grain boundary between G1 and G3 at LP0. The prior G1-G3 GB is an ATGB. (b) The propagation of the twin transforms the twin-G3 GB into a (nearly) STGB. (c) The partially-relaxed ATGB structure between G1-G3 prior to equilibration (which nucleates a twin). (d) A fully relaxed GB with identical grain boundary character to the twin-G3 GB. White, blue and black indicate hcp, fcc and boundary atoms in (a) and (b), whereas orange indicates hcp atoms in (c) and (d).}
    \label{fig:initGB_vs_finGB}
\end{figure}

As discussed in the introduction, one of the main unresolved questions relating to hcp twinning is the prediction of the twin nucleation site at grain boundaries in polycrystals.
Many studies focus on the connection between the grain boundary misorientation and the tendency for twins to nucleate on the loaded grain boundary;
the grain boundary inclination and the transformed grain boundary character are generally not considered to be relevant.
However, our results suggest that both of these could be important factors.
Figure \ref{fig:initGB_vs_finGB}a-b compares the G1-G3 grain boundaries before and after the nucleation of the twin.
Figure \ref{fig:initGB_vs_finGB}a shows that at LP0 the GB contains the prior G1-G3 boundary with ATGB character in the middle and transformed segments on both ends.
By comparison, the fully transformed GB shown in Fig.\ref{fig:initGB_vs_finGB}b has a nearly STGB character where the basal planes of the twin and G3 intersect the GB at similar angles.
However, the transformed GB has a corrugated topography with a structure that deviates from an STGB particularly at the intersections with stacking faults, suggesting the presence of defects with Burgers vector content there.

The transformation of the GB from a high-energy ATGB into a low-energy STGB during twinning significantly reduces the overall interfacial energy cost of nucleating the twin, as can be seen by comparing the green and orange curves in Fig.\ \ref{fig:regression2}.
Both the first and second regression analyses suggest a reduction in the grain boundary energy of roughly $\gamma_{\Delta\Gamma}\approx-0.0088\ \text{eV/\AA}^2$ ($-0.140\ \mathrm{J/m^2}$).
Two auxiliary MD simulations were conducted to verify that this value of $\gamma_{\Delta \Gamma}$ is plausible.
First, a copy of the original simulation cell containing the original G1-G3 GB was generated and relaxed by energy minimization.
Second, G1 was reoriented to the $\{10\overline{1}2\}$ twinned orientation while keeping the other grain orientations fixed, and the resulting simulation cell was equilibrated.
Representative slices of material around the GB regions were taken from both simulations and are shown in Fig.\ref{fig:initGB_vs_finGB}c-d.
The grain boundary energies of the G1-G3 and twinned G1-G3 boundaries are computed as:
\begin{equation}\label{eq:gb_energy}
    \gamma = \frac{E_\mathrm{tot} - NE_\mathrm{eq}}{A_\mathrm{GB}}
\end{equation}
where $E_\mathrm{tot}$ is the total potential energy of the slice, $N$ is the number of atoms in the slice, $E_\mathrm{eq}$ is the equilibrium per atom potential energy for bulk Mg, and $A_\mathrm{GB}$ is the area of the grain boundary.
The slices contain $16\ 520$ and $1388$ atoms respectively, and a value of $E_\mathrm{eq} = -1.51\ \mathrm{eV/atom}$ as measured by an MD simulation of a single crystal is consistent with the literature \cite{WuCurtinPotent2015}.
The computed energies for the initial (original) and final (twinned) grain boundaries are $\gamma_{i} = 0.0345 \ \text{eV/\AA}^2$ ($0.553 \ \mathrm{J/m}^2$) and $\gamma_{f}=0.077 \ \text{eV/\AA}^2$ ($0.124 \ \mathrm{J/m}^2$) respectively, where $\gamma_{f}$ is expected to be low since the transformed boundary is nearly a $\{10\overline{1}3\}$ twin boundary with the neighboring grains symmetrically tilted by $31.99^{\circ}$ with respect to the $[\overline{1}\overline{1}20]$ axis.
This is shown in Fig.\ \ref{fig:initGB_vs_finGB}d where the angles the basal planes of both grains make with the grain boundary are very close to $32^{\circ}$, with slight unequal deviations on each side.
Wang and Beyerlein \cite{WangBeyerlein2012} calculated the energy of this boundary to be $0.105 \ \mathrm{J/m}^2$, close to $\gamma_{f}$ found for the present structure.

Using these results, the reduction in the grain boundary energy could be as much as $\gamma_{f} - \gamma_{i} = -0.429 \ \mathrm{J/m}^2$, three times higher than the value $\gamma_{\Delta \Gamma}$ estimated from the regression analyses.
Since the actual transformation that takes place upon twinning does not yield a fully relaxed and equilibrated grain boundary (Fig.\ \ref{fig:initGB_vs_finGB}d) but rather a corrugated one with excess stresses (Fig.\ \ref{fig:initGB_vs_finGB}b), this is interpreted as indicating that the value of $\gamma_{\Delta \Gamma}$ from the regression analysis is physically reasonable.

It is significant that the original ATGB shown in Fig.\ \ref{fig:initGB_vs_finGB}c partially dissociated into the transformed GB shown in Fig.\ \ref{fig:initGB_vs_finGB}d when equilibrated at higher temperatures for longer durations, even in the absence of external strain.
This reaffirms that the change in the character of the prior GB can significantly reduce the energy barrier to twin nucleation.
This is consistent with the conclusions of Barrett et.\ al.\ \cite{Barrett2014,BarrettActa2014} who investigated twinning along a BP boundary in a bicrystal and found that the defect structure of ATGBs could be an important factor.

The implication is that the inclination of the prior grain boundary could be a possible controlling variable as well.
As evidence of this, unstable twins nucleated only on the upper of the two G1-G3 GBs (G1 shares two boundaries with G3 of the same misorientation in Fig.\ \ref{fig:simbox}) for which the inclination is such that $\{10\overline{1}2\}$ twinning yields an STGB.
Thus, the upper G1-G3 GB is a more favorable twinning location.
Our conclusion is that the change in the character of the prior GB should be considered in future twinning models.

\section{Conclusions}
\label{sec:conclusions}

Molecular dynamics simulations were performed to investigate the process by which a twin nucleates from a grain boundary in a columnar polycrystal.
A far-field strain was applied to drive the twin nucleation, but the potential energy of the twin as a function of time was measured with respect to the unstrained configuration as a consistent point of reference.
A model for the potential energy of the nucleating and growing twin was proposed that involves various geometric variables, parameters taken from the literature, and several unknown quantities estimated by linear regression.
Our main conclusions based on this energy model are as follows:
\begin{enumerate}
    \item The potential energy of the twin could not be satisfactorily modeled as a sum of contributions from only the twin boundary (comprised of coherent twin and basal-prismatic/prismatic-basal facets), stacking faults inside the twin, and elastic strain energy as estimated using the Eshelby inhomogeneity theory.
    That is, the potential energy of the nucleating twin necessarily involved more quantities (see below) than just those associated with visibly growing geometric structures in Fig.\ \ref{fig:summary_fig}.
    \item The reduction in grain boundary energy associated with the change in grain boundary character upon twinning can be substantial, and even to the point of offsetting the energy cost of the twin boundary and stacking fault creation.
    Specifically for the observed twinning event, the prior boundary decomposed from a high-energy ATGB structure into a low energy STGB structure with a $\{10\overline{1}3\}$ twin boundary relation.
    The other boundary in the periodic simulation cell with the same misorientation relation but a different inclination was inactive, indicating that grain boundary inclination is potentially a significant variable.
    Overall, this suggests that the change in grain boundary character during twinning could be a critical consideration when predicting twin nucleation sites.
    \item The elastic strain energy estimated using the Eshelby inhomogenity theory does not seem to be sufficient to explain the part of the twin potential energy that is approximately proportional to the volume of the twin.
    It is proposed that the stacking faults inside the twin are bounded by line defects with Burgers vector content embedded in the twin boundary and the prior grain boundary.
    As part of the extrinsic defect content of the boundaries, these line defects can have interacting elastic strain fields.
    While not conclusively establishing the existence of such interactions, including an energy contribution proportional to the logarithm of the normalized separation of the proposed line defects is sufficient for the Eshelby inhomogeneity theory to account for the part of the potential energy that is proportional to the twin volume.
\end{enumerate}
These conclusions are envisioned as part of an ongoing effort by the community to establish a micromechanical model which can be used to predict the activation energies and critical geometries of nucleating twins as functions of microstructure variables and far-field loading conditions.
It is true that both the molecular dynamics simulations and the energy model for twin formation developed here are restricted to thin films of $[\overline{1}\overline{1}20]$ fiber textured Mg, but we expect that the insights derived through the use of this simplified geometry will hold more generally, and particularly when the energy model is eventually extended to three dimensions.

\section*{Acknowledgements}
Authors acknowledge the support of the TUBITAK BIDEB 2214-A research program under the project code 1059B141900441.

\section*{Data availability}
The raw/processed data required to reproduce these findings cannot be shared at this time as the data also forms part of an ongoing study.
\bibliography{refs}

\begin{thebibliography}{10}
\expandafter\ifx\csname url\endcsname\relax
  \def\url#1{\texttt{#1}}\fi
\expandafter\ifx\csname urlprefix\endcsname\relax\def\urlprefix{URL }\fi
\expandafter\ifx\csname href\endcsname\relax
  \def\href#1#2{#2} \def\path#1{#1}\fi

\bibitem{JOMAgnew2004}
S.~R. Agnew, {Wrought magnesium: A 21st century outlook}, JOM 56~(5) (2004)
  20--21.
\newblock \href {https://doi.org/10.1007/s11837-004-0120-8}
  {\path{doi:10.1007/s11837-004-0120-8}}.

\bibitem{Agnew2005}
S.~R. Agnew, {\"{O}}.~Duygulu, {Plastic anisotropy and the role of non-basal
  slip in magnesium alloy AZ31B}, International Journal of Plasticity 21~(6)
  (2005) 1161--1193.
\newblock \href {https://doi.org/10.1016/j.ijplas.2004.05.018}
  {\path{doi:10.1016/j.ijplas.2004.05.018}}.

\bibitem{ArulKumar2017}
M.~{Arul Kumar}, I.~J. Beyerlein, C.~N. Tom{\'{e}}, {A measure of plastic
  anisotropy for hexagonal close packed metals: Application to alloying effects
  on the formability of Mg}, Journal of Alloys and Compounds 695 (2017)
  1488--1497.
\newblock \href {https://doi.org/10.1016/j.jallcom.2016.10.287}
  {\path{doi:10.1016/j.jallcom.2016.10.287}}.

\bibitem{Nie2020}
J.~F. Nie, K.~S. Shin, Z.~R. Zeng, {Microstructure, Deformation, and Property
  of Wrought Magnesium Alloys}, Metallurgical and Materials Transactions A:
  Physical Metallurgy and Materials Science 51~(12) (2020) 6045--6109.
\newblock \href {https://doi.org/10.1007/s11661-020-05974-z}
  {\path{doi:10.1007/s11661-020-05974-z}}.

\bibitem{Koike2005}
J.~Koike, {Enhanced deformation mechanisms by anisotropic plasticity in
  polycrystalline Mg alloys at room temperature}, Metallurgical and Materials
  Transactions A: Physical Metallurgy and Materials Science 36~(7) (2005)
  1689--1696.
\newblock \href {https://doi.org/10.1007/s11661-005-0032-4}
  {\path{doi:10.1007/s11661-005-0032-4}}.

\bibitem{WuNature2015}
Z.~Wu, W.~A. Curtin, {The origins of high hardening and low ductility in
  magnesium}, Nature 526~(7571) (2015) 62--67.
\newblock \href {https://doi.org/10.1038/nature15364}
  {\path{doi:10.1038/nature15364}}.

\bibitem{Aydiner2014}
C.~C. Aydiner, M.~A. Telemez, {Multiscale deformation heterogeneity in twinning
  magnesium investigated with in situ image correlation}, International Journal
  of Plasticity 56 (2014) 203--218.
\newblock \href {https://doi.org/10.1016/j.ijplas.2013.12.001}
  {\path{doi:10.1016/j.ijplas.2013.12.001}}.

\bibitem{Kapan2017}
E.~Kapan, N.~Shafaghi, C.~C. U\c{c}ar, Sevin\c{c}~Aydıner, {Texture-dependent
  character of strain heterogeneity in Magnesium AZ31 under reversed loading},
  Materials Science and Engineering A 684~(August 2016) (2017) 706--711.
\newblock \href {https://doi.org/10.1016/j.msea.2016.12.085}
  {\path{doi:10.1016/j.msea.2016.12.085}}.

\bibitem{Shafaghi2020}
N.~Shafaghi, E.~Kapan, C.~C. Aydıner, {Cyclic Strain Heterogeneity and Damage
  Formation in Rolled Magnesium Via In Situ Microscopic Image Correlation},
  Experimental Mechanics 60~(6) (2020) 735--751.
\newblock \href {https://doi.org/10.1007/s11340-020-00612-6}
  {\path{doi:10.1007/s11340-020-00612-6}}.

\bibitem{WangRaabe2014}
F.~Wang, S.~Sandl{\"{o}}bes, M.~Diehl, L.~Sharma, F.~Roters, D.~Raabe, {In situ
  observation of collective grain-scale mechanics in Mg and Mg-rare earth
  alloys}, Acta Materialia 80 (2014) 77--93.
\newblock \href {https://doi.org/10.1016/j.actamat.2014.07.048}
  {\path{doi:10.1016/j.actamat.2014.07.048}}.

\bibitem{Ucel2019}
I.~B. {\"{U}}{\c{c}}el, E.~Kapan, O.~T{\"{u}}rkoǧlu, C.~C. Aydiner, {In situ
  investigation of strain heterogeneity and microstructural shear bands in
  rolled Magnesium AZ31}, International Journal of Plasticity 118 (2019)
  233--251.
\newblock \href {https://doi.org/10.1016/j.ijplas.2019.02.008}
  {\path{doi:10.1016/j.ijplas.2019.02.008}}.

\bibitem{Orozco-Caballero2017}
A.~Orozco-Caballero, D.~Lunt, J.~D. Robson, J.~{Quinta da Fonseca}, {How
  magnesium accommodates local deformation incompatibility: A high-resolution
  digital image correlation study}, Acta Materialia 133 (2017) 367--379.
\newblock \href {https://doi.org/10.1016/j.actamat.2017.05.040}
  {\path{doi:10.1016/j.actamat.2017.05.040}}.

\bibitem{Yoo1991}
M.~H. Yoo, J.~K. Lee, {Deformation twinning in h.c.p. metals and alloys},
  Philosophical Magazine A: Physics of Condensed Matter, Structure, Defects and
  Mechanical Properties 63~(5) (1991) 987--1000.
\newblock \href {https://doi.org/10.1080/01418619108213931}
  {\path{doi:10.1080/01418619108213931}}.

\bibitem{Christian1995}
J.~W. Christian, S.~Mahajan, {Deformation Twinning}, Progress in Materials
  Science 39 (1995) 1--157.

\bibitem{Capolungo2008}
L.~Capolungo, I.~J. Beyerlein, {Nucleation and stability of twins in hcp
  metals}, Physical Review B - Condensed Matter and Materials Physics 78~(2)
  (2008) 1--19.
\newblock \href {https://doi.org/10.1103/PhysRevB.78.024117}
  {\path{doi:10.1103/PhysRevB.78.024117}}.

\bibitem{WangMRLett2013}
J.~Wang, S.~K. Yadav, J.~P. Hirth, C.~N. Tom{\'{e}}, I.~J. Beyerlein,
  {Pure-Shuffle Nucleation of Deformation Twins in Hexagonal-Close-Packed
  Metals}, Materials Research Letters 1~(3) (2013) 126--132.
\newblock \href {https://doi.org/10.1080/21663831.2013.792019}
  {\path{doi:10.1080/21663831.2013.792019}}.

\bibitem{BeyerleinJMPS2011}
I.~J. Beyerlein, R.~J. McCabe, C.~N. Tom{\'{e}}, {Effect of microstructure on
  the nucleation of deformation twins in polycrystalline high-purity magnesium:
  A multi-scale modeling study}, Journal of the Mechanics and Physics of Solids
  59~(5) (2011) 988--1003.
\newblock \href {https://doi.org/10.1016/j.jmps.2011.02.007}
  {\path{doi:10.1016/j.jmps.2011.02.007}}.

\bibitem{PaudelReview2021}
Y.~Paudel, D.~Giri, M.~W. Priddy, C.~D. Barrett, K.~Inal, M.~A. Tschopp,
  H.~Rhee, H.~{El Kadiri}, {A review on capturing twin nucleation in crystal
  plasticity for hexagonal metals}, Metals 11~(9) (2021) 1--53.
\newblock \href {https://doi.org/10.3390/met11091373}
  {\path{doi:10.3390/met11091373}}.

\bibitem{Paudel2020}
Y.~R. Paudel, C.~D. Barrett, H.~{El Kadiri}, {Full-Field Crystal Plasticity
  Modeling of $\{10\overline{1}2\}$ Twin Nucleation}, Minerals, Metals and
  Materials Series (2020) 141--146\href
  {https://doi.org/10.1007/978-3-030-36647-6-23}
  {\path{doi:10.1007/978-3-030-36647-6-23}}.

\bibitem{Roters2010}
F.~Roters, P.~Eisenlohr, L.~Hantcherli, D.~D. Tjahjanto, T.~R. Bieler,
  D.~Raabe, {Overview of constitutive laws, kinematics, homogenization and
  multiscale methods in crystal plasticity finite-element modeling: Theory,
  experiments, applications}, Acta Materialia 58~(4) (2010) 1152--1211.
\newblock \href {https://doi.org/10.1016/j.actamat.2009.10.058}
  {\path{doi:10.1016/j.actamat.2009.10.058}}.

\bibitem{ChengGhosh2017}
J.~Cheng, S.~Ghosh, {Crystal plasticity finite element modeling of discrete
  twin evolution in polycrystalline magnesium}, Journal of the Mechanics and
  Physics of Solids 99~(July 2016) (2017) 512--538.
\newblock \href {https://doi.org/10.1016/j.jmps.2016.12.008}
  {\path{doi:10.1016/j.jmps.2016.12.008}}.

\bibitem{ChengGhosh2018}
J.~Cheng, J.~Shen, R.~K. Mishra, S.~Ghosh, {Discrete twin evolution in Mg
  alloys using a novel crystal plasticity finite element model}, Acta
  Materialia 149 (2018) 142--153.
\newblock \href {https://doi.org/10.1016/j.actamat.2018.02.032}
  {\path{doi:10.1016/j.actamat.2018.02.032}}.

\bibitem{Kondo2014}
R.~Kondo, Y.~Tadano, K.~Shizawa, {A phase-field model of twinning and
  detwinning coupled with dislocation-based crystal plasticity for HCP metals},
  Computational Materials Science 95 (2014) 672--683.
\newblock \href {https://doi.org/10.1016/j.commatsci.2014.08.034}
  {\path{doi:10.1016/j.commatsci.2014.08.034}}.

\bibitem{Liu2018}
C.~Liu, P.~Shanthraj, M.~Diehl, F.~Roters, S.~Dong, J.~Dong, W.~Ding, D.~Raabe,
  {An integrated crystal plasticity-phase field model for spatially resolved
  twin nucleation, propagation, and growth in hexagonal materials},
  International Journal of Plasticity 106~(December 2017) (2018) 203--227.
\newblock \href {https://doi.org/10.1016/j.ijplas.2018.03.009}
  {\path{doi:10.1016/j.ijplas.2018.03.009}}.

\bibitem{Liu2021}
G.~Liu, H.~Mo, J.~Wang, Y.~Shen, {Coupled crystal plasticity finite
  element-phase field model with kinetics-controlled twinning mechanism for
  hexagonal metals}, Acta Materialia 202 (2021) 399--416.
\newblock \href {https://doi.org/10.1016/j.actamat.2020.11.002}
  {\path{doi:10.1016/j.actamat.2020.11.002}}.

\bibitem{WangLiu2013}
J.~Wang, L.~Liu, C.~N. Tom{\'{e}}, S.~X. Mao, S.~K. Gong, {Twinning and
  de-twinning via glide and climb of twinning dislocations along serrated
  coherent twin boundaries in hexagonal-close-packed metals}, Materials
  Research Letters 1~(2) (2013) 81--88.
\newblock \href {https://doi.org/10.1080/21663831.2013.779601}
  {\path{doi:10.1080/21663831.2013.779601}}.

\bibitem{Barrett2014}
C.~D. Barrett, H.~{El Kadiri}, {The roles of grain boundary dislocations and
  disclinations in the nucleation of $\{10\overline{1}2\}$ twinning}, Acta
  Materialia 63 (2014) 1--15.
\newblock \href {https://doi.org/10.1016/j.actamat.2013.09.012}
  {\path{doi:10.1016/j.actamat.2013.09.012}}.

\bibitem{GongMatResLett2017}
M.~Gong, J.~P. Hirth, Y.~Liu, Y.~Shen, J.~Wang, {Interface structures and
  twinning mechanisms of twins in hexagonal metals}, Materials Research Letters
  5~(7) (2017) 449--464.
\newblock \href {https://doi.org/10.1080/21663831.2017.1336496}
  {\path{doi:10.1080/21663831.2017.1336496}}.

\bibitem{Sun2014}
Q.~Sun, X.~Y. Zhang, Y.~Ren, J.~Tu, Q.~Liu, {Interfacial structure of {101¯2}
  twin tip in deformed magnesium alloy}, Scripta Materialia 90~(1) (2014)
  41--44.
\newblock \href {https://doi.org/10.1016/j.scriptamat.2014.07.012}
  {\path{doi:10.1016/j.scriptamat.2014.07.012}}.

\bibitem{Jiang2022}
L.~Jiang, M.~Gong, J.~Wang, Z.~Pan, X.~Wang, D.~Zhang, Y.~M. Wang, J.~Ciston,
  A.~M. Minor, M.~Xu, X.~Pan, T.~J. Rupert, S.~Mahajan, E.~J. Lavernia, I.~J.
  Beyerlein, J.~M. Schoenung, {Visualization and validation of twin nucleation
  and early-stage growth in magnesium}, Nature Communications 13~(1) (2022)
  1--11.
\newblock \href {https://doi.org/10.1038/s41467-021-27591-z}
  {\path{doi:10.1038/s41467-021-27591-z}}.

\bibitem{Ostapovets2014}
A.~Ostapovets, A.~Serra, {Characterization of the matrix – twin interface of
  a $(10\overline{1}2)$ twin during growth}, Philosophical Magazine 94~(25)
  (2014) 2827--2839.
\newblock \href {https://doi.org/10.1080/14786435.2014.933906}
  {\path{doi:10.1080/14786435.2014.933906}}.

\bibitem{ElKadiri2015}
H.~{El Kadiri}, C.~D. Barrett, J.~Wang, C.~N. Tom{\'{e}}, {Why are
  $\{10\overline{1}2\}$ twins profuse in magnesium?}, Acta Materialia 85 (2015)
  354--361.
\newblock \href {https://doi.org/10.1016/j.actamat.2014.11.033}
  {\path{doi:10.1016/j.actamat.2014.11.033}}.

\bibitem{Liu2016}
Y.~Liu, N.~Li, S.~Shao, M.~Gong, J.~Wang, R.~J. McCabe, Y.~Jiang, C.~N.
  Tom{\'{e}}, {Characterizing the boundary lateral to the shear direction of
  deformation twins in magnesium}, Nature Communications 7 (2016) 8--13.
\newblock \href {https://doi.org/10.1038/ncomms11577}
  {\path{doi:10.1038/ncomms11577}}.

\bibitem{WangScience2020}
S.~Wang, M.~Gong, R.~J. McCabe, L.~Capolungo, J.~Wang, C.~N. Tom{\'{e}},
  {Characteristic boundaries associated with three-dimensional twins in
  hexagonal metals}, Science Advances 6~(28) (2020).
\newblock \href {https://doi.org/10.1126/sciadv.aaz2600}
  {\path{doi:10.1126/sciadv.aaz2600}}.

\bibitem{GongActa2018}
M.~Gong, G.~Liu, J.~Wang, L.~Capolungo, C.~N. Tom{\'{e}}, {Atomistic
  simulations of interaction between basal dislocations and three-dimensional
  twins in magnesium}, Acta Materialia 155 (2018) 187--198.
\newblock \href {https://doi.org/10.1016/j.actamat.2018.05.066}
  {\path{doi:10.1016/j.actamat.2018.05.066}}.

\bibitem{GongActa2021}
M.~Gong, J.~Graham, V.~Taupin, L.~Capolungo, {The effects of stress,
  temperature and facet structure on growth of $\{10\overline{1}2\}$ twins in
  Mg: A molecular dynamics and phase field study}, Acta Materialia 208 (2021).
\newblock \href {https://doi.org/10.1016/j.actamat.2020.116603}
  {\path{doi:10.1016/j.actamat.2020.116603}}.

\bibitem{Spearot2020}
D.~E. Spearot, V.~Taupin, K.~Dang, L.~Capolungo, {Structure and kinetics of
  three-dimensional defects on the {101¯2} twin boundary in magnesium:
  Atomistic and phase-field simulations}, Mechanics of Materials 143 (2020).
\newblock \href {https://doi.org/10.1016/j.mechmat.2020.103314}
  {\path{doi:10.1016/j.mechmat.2020.103314}}.

\bibitem{BeyerleinPhilMag2010}
I.~J. Beyerlein, L.~Capolungo, P.~E. Marshall, R.~J. McCabe, C.~N. Tome,
  {Statistical analyses of deformation twinning in magnesium}, Philosophical
  Magazine 90~(16) (2010) 2161--2190.
\newblock \href {https://doi.org/10.1080/14786431003630835}
  {\path{doi:10.1080/14786431003630835}}.

\bibitem{Khosravani2015}
A.~Khosravani, D.~T. Fullwood, B.~L. Adams, T.~M. Rampton, M.~P. Miles, R.~K.
  Mishra, {Nucleation and propagation of $\{10\overline{1}2\}$ twins in AZ31
  magnesium alloy}, Acta Materialia 100~(January 2018) (2015) 202--214.
\newblock \href {https://doi.org/10.1016/j.actamat.2015.08.024}
  {\path{doi:10.1016/j.actamat.2015.08.024}}.

\bibitem{WangScripta2010}
J.~Wang, I.~J. Beyerlein, C.~N. Tom{\'{e}}, {An atomic and probabilistic
  perspective on twin nucleation in Mg}, Scripta Materialia 63~(7) (2010)
  741--746.
\newblock \href {https://doi.org/10.1016/j.scriptamat.2010.01.047}
  {\path{doi:10.1016/j.scriptamat.2010.01.047}}.

\bibitem{WangIntJPlas2013}
J.~Wang, I.~J. Beyerlein, C.~N. Tom{\'{e}}, {Reactions of lattice dislocations
  with grain boundaries in Mg: Implications on the micro scale from
  atomic-scale calculations}, International Journal of Plasticity 56 (2013)
  156--172.
\newblock \href {https://doi.org/10.1016/j.ijplas.2013.11.009}
  {\path{doi:10.1016/j.ijplas.2013.11.009}}.

\bibitem{BarrettActa2014}
C.~D. Barrett, H.~{El Kadiri}, {Impact of deformation faceting on
  $\{10\overline{1}2\}$, $\{10\overline{1}1\}$ and $\{10\overline{1}3\}$
  embryonic twin nucleation in hexagonal close-packed metals}, Acta Materialia
  70 (2014) 137--161.
\newblock \href {https://doi.org/10.1016/j.actamat.2014.02.018}
  {\path{doi:10.1016/j.actamat.2014.02.018}}.

\bibitem{Giri2020}
D.~Giri, H.~ElKadiri, K.~R. Limmer, C.~D. Barrett, {An atomistic gateway into
  capturing twin nucleation in crystal plasticity}, Philosophical Magazine
  Letters 100~(8) (2020) 375--385.
\newblock \href {https://doi.org/10.1080/09500839.2020.1774932}
  {\path{doi:10.1080/09500839.2020.1774932}}.

\bibitem{Kim2010}
D.~H. Kim, M.~V. Manuel, F.~Ebrahimi, J.~S. Tulenko, S.~R. Phillpot,
  {Deformation processes in [11-20]-textured nanocrystalline Mg by molecular
  dynamics simulation}, Acta Materialia 58~(19) (2010) 6217--6229.
\newblock \href {https://doi.org/10.1016/j.actamat.2010.07.036}
  {\path{doi:10.1016/j.actamat.2010.07.036}}.

\bibitem{Miyazawa2015}
N.~Miyazawa, T.~Yoshida, M.~Yuasa, Y.~Chino, M.~Mabuchi, {Effect of segregated
  Al on {1012} and {1011} twinning in Mg}, Journal of Materials Research
  30~(23) (2015) 3629--3641.
\newblock \href {https://doi.org/10.1557/jmr.2015.330}
  {\path{doi:10.1557/jmr.2015.330}}.

\bibitem{Agarwal2019}
G.~Agarwal, A.~M. Dongare, {Deformation Twinning in Polycrystalline Mg
  Microstructures at High Strain Rates at the Atomic Scales}, Scientific
  Reports 9~(August 2018) (2019) 1--11.
\newblock \href {https://doi.org/10.1038/s41598-019-39958-w}
  {\path{doi:10.1038/s41598-019-39958-w}}.

\bibitem{Wang2019}
F.~Wang, C.~D. Barrett, R.~J. McCabe, H.~{El Kadiri}, L.~Capolungo, S.~R.
  Agnew, \href{https://doi.org/10.1016/j.actamat.2018.12.003}{{Dislocation
  induced twin growth and formation of basal stacking faults in
  $\{101\overline{2}\}$ twins in pure Mg}}, Acta Materialia 165 (2019)
  471--485.
\newblock \href {https://doi.org/10.1016/j.actamat.2018.12.003}
  {\path{doi:10.1016/j.actamat.2018.12.003}}.
\newline\urlprefix\url{https://doi.org/10.1016/j.actamat.2018.12.003}

\bibitem{Zhou2021}
X.~Zhou, H.~Su, H.~Ye, Z.~Yang,
  \href{https://doi.org/10.1016/j.actamat.2021.117170}{{Removing
  basal-dissociated $\langle c + a \rangle$ dislocations by
  $\{101\overline{2}\}$ deformation twinning in magnesium alloys}}, Acta
  Materialia 217 (2021) 117170.
\newblock \href {https://doi.org/10.1016/j.actamat.2021.117170}
  {\path{doi:10.1016/j.actamat.2021.117170}}.
\newline\urlprefix\url{https://doi.org/10.1016/j.actamat.2021.117170}

\bibitem{Yue2022}
Y.~Yue, Y.~Zhang, J.~F. Nie, {Stability of single-atomic-layer-height
  disconnections on (101¯2) twin boundary in Mg}, Scripta Materialia 209
  (2022) 114407.
\newblock \href {https://doi.org/10.1016/j.scriptamat.2021.114407}
  {\path{doi:10.1016/j.scriptamat.2021.114407}}.

\bibitem{LAMMPS}
A.~P. Thompson, H.~M. Aktulga, R.~Berger, D.~S. Bolintineanu, W.~M. Brown,
  P.~S. Crozier, P.~J. in~'t Veld, A.~Kohlmeyer, S.~G. Moore, T.~D. Nguyen,
  R.~Shan, M.~J. Stevens, J.~Tranchida, C.~Trott, S.~J. Plimpton, {LAMMPS} - a
  flexible simulation tool for particle-based materials modeling at the atomic,
  meso, and continuum scales, Comp. Phys. Comm. 271 (2022) 108171.

\bibitem{WuCurtinPotent2015}
Z.~Wu, M.~F. Francis, W.~A. Curtin, {Magnesium interatomic potential for
  simulating plasticity and fracture phenomena}, Modelling and Simulation in
  Materials Science and Engineering Modelling 23 (2015) 1--19.
\newblock \href {https://doi.org/10.1088/0965-0393/23/1/015004}
  {\path{doi:10.1088/0965-0393/23/1/015004}}.

\bibitem{ATOMSK}
P.~Hirel, {Atomsk: A tool for manipulating and converting atomic data files},
  Computer Physics Communications 197 (2015) 212--219.
\newblock \href {https://doi.org/10.1016/j.cpc.2015.07.012}
  {\path{doi:10.1016/j.cpc.2015.07.012}}.

\bibitem{Shinoda2004}
W.~Shinoda, M.~Shiga, M.~Mikami, {Rapid estimation of elastic constants by
  molecular dynamics simulation under constant stress}, Physical Review B -
  Condensed Matter and Materials Physics 69~(13) (2004) 16--18.
\newblock \href {https://doi.org/10.1103/PhysRevB.69.134103}
  {\path{doi:10.1103/PhysRevB.69.134103}}.

\bibitem{Stukowski2010}
A.~Stukowski, {Visualization and analysis of atomistic simulation data with
  OVITO-the Open Visualization Tool}, Modelling and Simulation in Materials
  Science and Engineering 18~(1) (2010).
\newblock \href {https://doi.org/10.1088/0965-0393/18/1/015012}
  {\path{doi:10.1088/0965-0393/18/1/015012}}.

\bibitem{Eshelby1957}
J.~D. Eshelby, {The determination of the elastic field of an ellipsoidal
  inclusion, and related problems}, Proceedings of the royal society of London.
  Series A. Mathematical and physical sciences 241~(1226) (1957) 376--396.
\newblock \href {https://doi.org/https://doi.org/10.1098/rspa.1957.0133}
  {\path{doi:https://doi.org/10.1098/rspa.1957.0133}}.

\bibitem{Pond2016}
R.~C. Pond, J.~P. Hirth, A.~Serra, D.~J. Bacon,
  \href{https://doi.org/10.1080/21663831.2016.1165298}{{Atomic displacements
  accompanying deformation twinning: shears and shuffles}}, Materials Research
  Letters 4~(4) (2016) 185--190.
\newblock \href {https://doi.org/10.1080/21663831.2016.1165298}
  {\path{doi:10.1080/21663831.2016.1165298}}.
\newline\urlprefix\url{https://doi.org/10.1080/21663831.2016.1165298}

\bibitem{Zhao2019}
X.~Zhao, H.~Chen, N.~Wilson, Q.~Liu, J.~F. Nie,
  \href{http://dx.doi.org/10.1038/s41467-019-10921-7}{{Direct observation and
  impact of co-segregated atoms in magnesium having multiple alloying
  elements}}, Nature Communications 10~(1) (2019) 1--7.
\newblock \href {https://doi.org/10.1038/s41467-019-10921-7}
  {\path{doi:10.1038/s41467-019-10921-7}}.
\newline\urlprefix\url{http://dx.doi.org/10.1038/s41467-019-10921-7}

\bibitem{WuActa2016}
Z.~Wu, B.~Yin, W.~A. Curtin, {Energetics of dislocation transformations in hcp
  metals}, Acta Materialia 119 (2016) 203--217.
\newblock \href {https://doi.org/10.1016/j.actamat.2016.08.002}
  {\path{doi:10.1016/j.actamat.2016.08.002}}.

\bibitem{DangActa2020}
K.~Dang, S.~Wang, M.~Gong, R.~J. McCabe, J.~Wang, L.~Capolungo,
  \href{https://doi.org/10.1016/j.actamat.2019.11.070}{{Formation and stability
  of long basal-prismatic facets in Mg}}, Acta Materialia 185 (2020) 119--128.
\newblock \href {https://doi.org/10.1016/j.actamat.2019.11.070}
  {\path{doi:10.1016/j.actamat.2019.11.070}}.
\newline\urlprefix\url{https://doi.org/10.1016/j.actamat.2019.11.070}

\bibitem{Barrett2014b}
C.~D. Barrett, H.~{El Kadiri}, {Impact of deformation faceting on
  $\{101\overline{2}\}$, $\{101\overline{1}\}$ and $\{101\overline{3}\}$
  embryonic twin nucleation in hexagonal close-packed metals}, Acta Materialia
  70 (2014) 137--161.
\newblock \href {https://doi.org/10.1016/j.actamat.2014.02.018}
  {\path{doi:10.1016/j.actamat.2014.02.018}}.

\bibitem{BarnettCaiJMPS2018}
D.~M. Barnett, W.~Cai, {Properties of the Eshelby tensor and existence of the
  equivalent ellipsoidal inclusion solution}, Journal of the Mechanics and
  Physics of Solids 121 (2018) 71--80.
\newblock \href {https://doi.org/10.1016/j.jmps.2018.07.019}
  {\path{doi:10.1016/j.jmps.2018.07.019}}.

\bibitem{mura2013micromechanics}
T.~Mura, Micromechanics of defects in solids, Springer Science \& Business
  Media, 2013.

\bibitem{CapolungoIsotropic2010}
L.~Capolungo, I.~J. Beyerlein, Z.~Qwang, {The role of elastic anisotropy on
  plasticity in hcp metals: A three-dimensional dislocation dynamics study},
  Modelling and Simulation in Materials Science and Engineering 18~(8) (2010).
\newblock \href {https://doi.org/10.1088/0965-0393/18/8/085002}
  {\path{doi:10.1088/0965-0393/18/8/085002}}.

\bibitem{TromansIsotropic2011}
D.~Tromans, Elastic anisotropy of hcp metal crystals and polycrystals, Int. J.
  Res. Rev. Appl. Sci. 6~(4) (2011) 462--483.

\bibitem{WangScripta2018}
F.~Wang, K.~Hazeli, K.~D. Molodov, C.~D. Barrett, T.~Al-Samman, D.~A. Molodov,
  A.~Kontsos, K.~T. Ramesh, H.~{El Kadiri}, S.~R. Agnew,
  \href{https://doi.org/10.1016/j.scriptamat.2017.09.015}{{Characteristic
  dislocation substructure in 101¯2 twins in hexagonal metals}}, Scripta
  Materialia 143 (2018) 81--85.
\newblock \href {https://doi.org/10.1016/j.scriptamat.2017.09.015}
  {\path{doi:10.1016/j.scriptamat.2017.09.015}}.
\newline\urlprefix\url{https://doi.org/10.1016/j.scriptamat.2017.09.015}

\bibitem{HeActa2020}
C.~He, Y.~Zhang, C.~Q. Liu, Y.~Yue, H.~W. Chen, J.~F. Nie, {Unexpected partial
  dislocations within stacking faults in a cold deformed Mg-Bi alloy}, Acta
  Materialia 188~(2020) (2020) 328--343.
\newblock \href {https://doi.org/10.1016/j.actamat.2020.02.010}
  {\path{doi:10.1016/j.actamat.2020.02.010}}.

\bibitem{Pond1999}
R.~C. Pond, A.~Serra, D.~J. Bacon, {Dislocations in interfaces in the h.c.p.
  metals - II. Mechanisms of defect mobility under stress}, Acta Materialia
  47~(5) (1999) 1441--1453.
\newblock \href {https://doi.org/10.1016/S1359-6454(99)00017-8}
  {\path{doi:10.1016/S1359-6454(99)00017-8}}.

\bibitem{WangBeyerlein2012}
J.~Wang, I.~J. Beyerlein, {Atomic structures of symmetric tilt grain boundaries
  in hexagonal close packed (hcp) crystals}, Modelling and Simulation in
  Materials Science and Engineering 20~(2) (2012) 0--22.
\newblock \href {https://doi.org/10.1088/0965-0393/20/2/024002}
  {\path{doi:10.1088/0965-0393/20/2/024002}}.

\end{thebibliography}

\end{document}